# Correlations and Semimetallic Behaviors in Pyrochlore Oxide $Cd_2Re_2O_7$

Zenji HIROI*, Masafumi HANAWA, Yuji MURAOKA and Hisatomo HARIMA[1]

*Institute for Solid State Physics, University of Tokyo, Kashiwa, Chiba 277-8581, Japan*
*[1]ISIR-Sanken, Osaka University, Ibaraki, Osaka 567-0047, Japan*

Electronic properties of the metallic pyrochlore oxide $Cd_2Re_2O_7$ are studied by means of electrical resistivity and Hall measurements. Semimetallic band structures are revealed as expected from band structure calculations. It is found that large changes in carrier density and mass occur at the structural phase transition at $T_{s1}$ = 200 K. A large mass enhancement is observed, particularly for the high-temperature phase with the ideal pyrochlore structure, suggesting that an anomalous correlation has an important effect on the itinerant electrons in the pyrochlore lattice.

*E-mail address: hiroi@issp.u-tokyo.ac.jp

Superconductivity below a critical temperature $T_c$ = 1.0 K was found recently in pyrochlore oxide $Cd_2Re_2O_7$.[1-3] The superconductivity is a conventional, being a weak-coupling Bardeen-Cooper-Schrieffer (BCS) type with ordinary superconducting parameters.[1, 4-6] However, unusual successive structural phase transitions have been found at $T_{s1}$ = 200 K[7, 8] and $T_{s2}$ = 120 K,[9] which apparently reflect an intriguing interplay between the crystal and electronic structures for the itinerant electron system on the pyrochlore lattice. Three phases identified are phase I above $T_{s1}$ with the ideal pyrochlore structure, phase II at the intermediate temperature range, and phase III below $T_{s2}$. The two low-temperature phases possess slightly distorted pyrochlore structures.[7, 10] Applying high pressure effectively stabilizes the high-temperature phases, and superconductivity seems to disappear above a critical pressure of 3.5 GPa where $T_{s1}$ drops to $T$ = 0.[11] Structural deformations seem to be necessary for the occurrence of superconductivity at low temperature.

According to recent band structure calculations by Harima[12] and Singh et al.,[13] phase I is a compensated semimetal with low carrier density: there are electron pockets at the $\Gamma$ point and hole pockets at the $K$ point in the momentum space. Here, we present experimental evidence of semimetallic band structures for phase I as well as phase III.

Single crystals of $Cd_2Re_2O_7$ prepared as reported previously[1, 4] were used for all measurements. Electrical resistivity $\rho$ was measured between 2 and 300 K in magnetic fields up to 14 T by the standard four-probe method in a Quantum Design PPMS system. Magnetoresistance was measured by rotating a sample in a magnetic field of 14 T applied along the [001], [110], and [111] directions of the cubic unit cell, while keeping as constant the current flow perpendicular to the field. Hall measurements were carried out using reversed field sweeps up to 14 T at constant temperatures between 5 and 260 K. Hall voltage $V_H$ was obtained from the even component of the transverse voltage. For the measurements we used a polished crystal with dimensions of 1.5 mm x 3.0 mm x 90 μm.

As reported previously, phase I above $T_{s1}$ = 200 K is a poor metal, while the low-temperature phases II and III are good metals.[1] The resistivity measured at a critical pressure of 3.5 GPa presents the intrinsic temperature dependence of resistivity for phase I, which is shown in Fig. 1.[11] By comparing the two resistivity curves for phases I and III at low temperature, it is found that the residual resistivity $\rho_0$ is one order larger in phase I than in phase III. Various electronic parameters for these phases are summarized in Table I. Jin et al. reported that the $\rho$ of $Cd_2Re_2O_7$ crystals at ambient pressure showed a quadratic temperature dependence in a wide range between 2 and 60 K.[3] However, our result consistently shows a much narrower temperature window ($T \leq 5$ K) for a $T^2$ behavior, and is approximately proportional to $T^3$ below 20 K. This is independent of the value of residual resistivity. Since the system is a low-carrier semimetal with small Fermi energy $E_F$ as will be discussed later, the temperature window for the quadratic behavior characteristic of Fermi liquid should be narrow ($k_B T \ll E_F$). The coefficient $A$ of the quadratic term for phase III was determined to be 4.0 x $10^{-3}$ μΩcm/K$^2$ using the data from the best crystal with the lowest $\rho_0$ of 11.5 μΩcm.[4] This $A$ value is much smaller than that reported by Jin et al. (2.4 x $10^{-2}$ μΩcm/K$^2$).[3] From a Sommerfeld coefficient $\gamma$ of 15.1 mJ/K$^2$ mol Re obtained by heat capacity measurements,[1] a Kadowaki-Woods ratio $A/\gamma^2$ is evaluated to be 1.75 x $10^{-5}$ μΩcm/(mJ/K mol Re)$^2$, which is close to the universal value of $a_0$ = 1.0 x $10^{-5}$ μΩcm/(mJ/K mol)$^2$ for heavy Fermion systems. On the other hand, as shown in the inset in Fig. 1, the $\rho$ of phase I exhibits an apparent quadratic temperature dependence in a wider temperature range with a much larger $A$ value of 4.5 x $10^{-2}$ μΩcm/K$^2$ than that of $\rho$ of phase III. This implies heavier carriers existing in phase I.

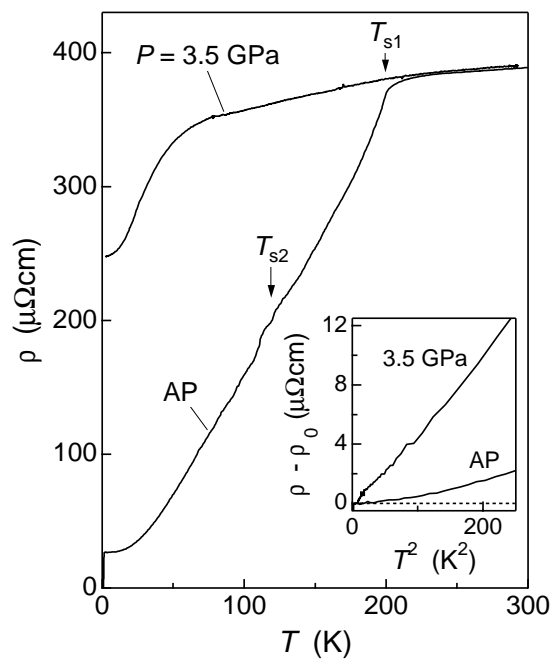

Fig. 1. Temperature dependence of electrical resistivity ($\rho$) measured in a $Cd_2Re_2O_7$ single crystal at ambient pressure (AP) and high pressure of 3.5 GPa. The latter presents an intrinsic behavior for the high-temperature phase I. $T_{s1}$ and $T_{s2}$ are structural transition temperatures. The inset shows the $\rho$-$\rho_0$ plot against $T^2$, where $\rho_0$ is the residual resistivity (26.7 μΩcm for AP and 247.4 μΩcm for 3.5 GPa).

Figure 2 shows the temperature dependence of Hall coefficient $R_H$. Hall voltage $V_H$ was very small and proportional to the magnitude of applied fields up to 14 T, as shown in the inset in Fig. 2. $R_H$ was determined from the slope at each temperature. It is small and positive around room temperature, changes its sign near 200 K, and becomes larger below 120 K with a negative sign. Then, its temperature dependence is saturated below 50 K with $R_H$ = -3.2 x $10^{-10}$ m$^3$/C at $T$ = 2 K. In contrast, Jin et al. reported a substantially different temperature dependence for $R_H$ with a much larger value of approximately -8 x $10^{-10}$ m$^3$/C at $T$ = 2 K.[8] It is noted that the two $R_H$ values previously measured by Sleight (marked by cross in Fig. 2) are in good agreement with our data. The obtained small $R_H$ and unusually large temperature dependence are actually consistent with such a semimetallic band structure as expected from the calculations.[12, 13]

We have measured the transverse magnetoresistance $\Delta\rho$ for Cd$_2$Re$_2$O$_7$ under magnetic fields up to 14 T. Figure 3(a) shows its field dependence along the three crystallographic directions measured at $T$ = 2 K. The magnetoresistance increases quadratically at first with increasing field, and further increases gradually up to $\mu_0 H$ = 14 T without any signs of saturation. In general, $\Delta\rho$ increases at a high magnetic field in the case of a band structure with an open Fermi surface, while it tends to be saturated in the case of a band structure with a closed Fermi surface. Since no saturations are detected along the three major directions in the present case, the possibility for open orbits must be small. Assuming a compensated semimetal gives an alternative explanation for such an increase in $\Delta\rho$ as observed here. This suggests that low-temperature phase III, as well as high-temperature phase I, is also a semimetal. Probably the small and round electron-like Fermi surface at the zone center predicted for phase I dominates the transport properties at low temperature.[12] The angular dependence of $\Delta\rho$ exhibits a small but significant anisotropy possibly reflecting the topology of the Fermi surface (Fig. 3(b)). It shows a marked valley when the field is parallel to the [110] direction. To be noted is that $\Delta\rho$ decreases steeply on heating and almost vanishes near the critical temperatures $T_{s1}$ and $T_{s2}$.[9] However, it recovers at intermediate temperatures between 120 and 200 K, namely, for phase II it is positive in [001], almost zero in [111], and negative in [110] directions. These temperature dependences indicate that marked changes in the Fermi surface take place successively at $T_{s1}$ and $T_{s2}$.

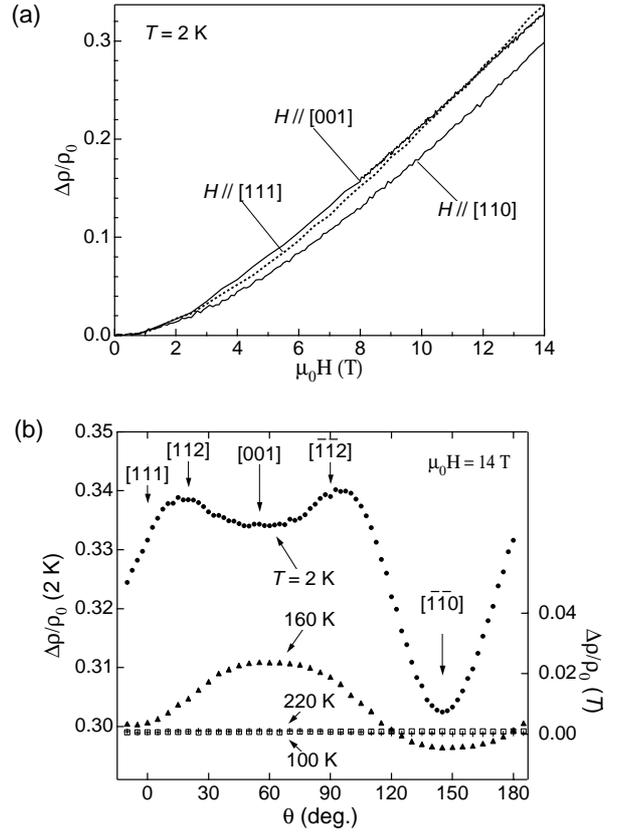

Fig. 3. (a) Magnetoresistance $\Delta\rho$ divided by $\rho_0$ (12.23 μΩcm) measured at $T$ = 2 K in magnetic fields applied along the [001], [110], and [111] directions. (b) Angle dependence of $\Delta\rho/\rho_0$ measured in a field $\mu_0 H$ = 14 T at $T$ = 2 K (left axis) and $T$ = 100 K, 160 K, and 220 K (right axis). The electrical current flowed along the [110] direction which was always perpendicular to the field.

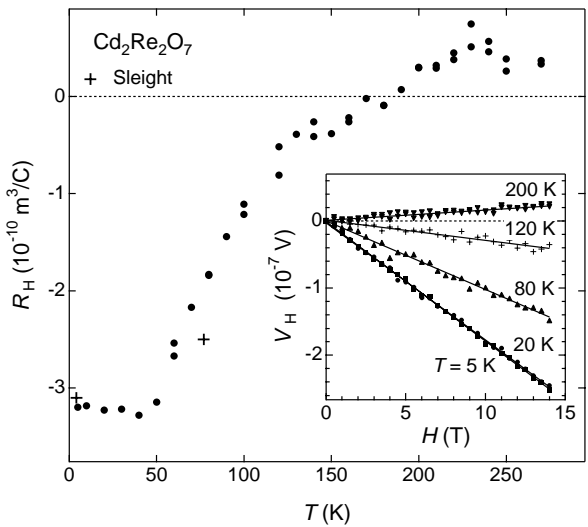

Fig. 2. Temperature dependence of Hall coefficient $R_H$. The two crosses present the data given by Sleight. The inset shows the field dependence of Hall voltage $V_H$.

We have estimated the carrier density of phase III assuming a compensated semimetal. For a compensated metal, resistivity $\rho$, Hall coefficient $R_H$, and magnetoresistance $\Delta\rho/\rho_0$ are given

by

$$\frac{1}{\rho} = ne(\mu_e + \mu_h) \rightarrow (1),$$

$$R_H = \frac{1}{ne}\frac{\mu_h - \mu_e}{\mu_e + \mu_h} \rightarrow (2),$$

$$\Delta\rho/\rho_0 = \mu_e\mu_h H^2 \rightarrow (3),$$

where $n$ is the carrier density of electrons or holes, and $\mu_e$ and $\mu_h$ are electron and hole mobilities, respectively. The last equation is valid in the high-field limit. From the present experimental data ($\rho = 11.5$ µΩcm, $R_H = -3.2 \times 10^{-10}$ m$^3$/C at $T = 2$ K, and $\Delta\rho/\rho_0 = 1.13 \times 10^5$ Vscm$^{-2}$ at $T = 5$ K), we obtained $n = 8.1 \times 10^{20}$ cm$^{-3}$ (0.055/Re), $\mu_e = 350$ cm$^2$V$^{-1}$s$^{-1}$, and $\mu_h = 323$ cm$^2$V$^{-1}$s$^{-1}$. This carrier density is considerably larger than that of conventional semimetals such as Sb ($n = 5.5 \times 10^{19}$ cm$^{-3}$) and Bi ($n = 2.8 \times 10^{17}$ cm$^{-3}$), while the mobility is small. The negative sign of $R_H$ observed at 2 K comes from the slightly larger electron mobility, and the much smaller positive $R_H$ above 200 K indicates that $\mu_e \sim \mu_h$. We have estimated also the mean-free-path $l$ at $T = 2$ K to be 81 nm using

$$\rho_0 = \frac{h(3\pi^2)^{\frac{1}{3}}}{e^2 l} n^{-\frac{2}{3}} \rightarrow (4).$$

With regard to the carrier density for phase I, the observed large residual resistivity under high pressure indicates that the carrier density in phase I is much smaller than that in phase III. Using eq. (1), by assuming the same values for mobility, $n = 3.6 \times 10^{19}$ cm$^{-3}$. On the other hand, band structure calculations, which were carried out by using the positional parameter of the 48$f$ oxygen ($x$) determined by means of powder neutron diffraction ($x = 0.31728$), yield $1.5 \times 10^{20}$ cm$^{-3}$.[14] In all cases the carrier density of phase I is markedly reduced compared with that of phase III (Table I). This is one of the reasons why the resistivity is reduced markedly at $T_{s1}$. The observed dramatic changes in the electronic structure imply a strong coupling between the crystal and electronic structures. This must be related to instability inherent to the itinerant electrons on the pyrochlore lattice. The band structure calculations for phase I revealed a semimetallic band structure with a Fermi surface consisting of two kinds of electron-like bands centered at the $\Gamma$ point and one kind of hole-like bands at the $K$ point, as schematically shown in Fig. 4.[12,13] To be noted is that unusually large (twelvefold) degeneracy exists for the latter, which must come from the high symmetry of the pyrochlore structure. As a result, the hole-like bands make a larger contribution to the density-of-state (DOS) (4.51 mJ/K$^2$ mol Re) than the electron-like bands (1.22 mJ/K$^2$ mol Re). The experimental result that phase I with lower carrier density exhibits a larger DOS than phase III implies that the degeneracy of the hole bands is lifted partially below the structural transition at $T_{s1}$ and a fraction of the hole bands sink below the Fermi level. Harima suggested that a certain lattice deformation would give rise to such a change in the band structure.[12] In this sense the transition at $T_{s1}$ can be regarded as a band Jahn-Teller transition.

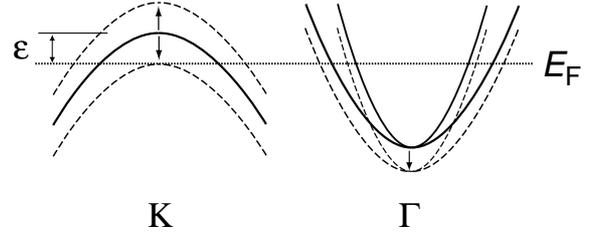

Fig. 4. Schematic representation of the semimetallic band structure for Cd$_2$Re$_2$O$_7$ above (solid lines) and below (broken lines) $T_{s1}$. On the basis of the band structure calculations, there are two kinds of electron-like bands at the $\Gamma$ point and one kind of hole-like bands at the $K$ point. The latter possesses twelvefold degeneracy, which may be lifted partially below $T_{s1}$ due to the band Jahn-Teller effect.

Figure 4 illustrates a simplified semimetallic band structure and its possible modification below $T_{s1}$. In general, when a conduction band has multivalleys, the energy of the top or bottom of the band from the Fermi energy, $\varepsilon$, is given by

$$\varepsilon = \frac{h^2}{2m_0}\left(3\pi^2 \frac{n}{v}\right)^{\frac{2}{3}},$$ where $v$ is the number of valleys.

Therefore, $n$ is proportional to $v\varepsilon^{\frac{3}{2}}$, while $\gamma$ (DOS) is proportional to $v\varepsilon^{\frac{1}{2}}$ ($\gamma \propto \frac{\partial n}{\partial \varepsilon}$). It is intuitive to consider how $n$ and $\gamma$ change, when the hole bands split into two due to the band Jahn-Teller effect. For example, in the case that half of the hole bands shift upward to $\varepsilon' = 2\varepsilon$ and the other half downward to $\varepsilon'' = 0$, $v' = v/2$. Such a change would increase $n$ by a factor of $\sqrt{2}$, while $\gamma$ would decrease by $1/\sqrt{2}$. Therefore, the experimental observations are explained qualitatively by this band Jahn-Teller concept. Quantitative discussions require structural data below $T_{s1}$ and the following band structure calculations for phases II and III.

In order to understand the fundamental physics of the present compound, it is crucial to estimate the mass enhancement of carriers. First we discuss phase III. There are no band calculations to be compared with the experimental value for $\gamma$ (15.1 mJ/K$^2$ mol Re).[1] One measure for electron correlations is the Wilson ratio, $R_W = \frac{\pi^2}{3}\left(\frac{k_B}{\mu_B}\right)^2\left(\frac{\chi_s}{\gamma}\right)$, which is unity for free electrons and two for strongly correlated electrons. Using the experimental value of spin susceptibility $\chi_s$ (7.5 x $10^{-5}$ cm$^3$/mol Re), we obtained an unusually small value of 0.34 for $R_W$.[15] This indicates that the correlation here enhances mainly $\gamma$, and thus, it is not a conventional electron correlation. Electron-phonon interactions must have a minor contribution, because the superconductivity of $Cd_2Re_2O_7$ is understood well within the framework of the weak-coupling BCS theory. Vyaselev *et al.* reported that the Re relaxation rate is unexpectedly larger than the Cd one.[15] This suggests that a fluctuation relevant to Re 5d electrons is crucial. Since there are two electrons in the $t_{2g}$ orbital and the orbital degree of freedom may remain, it is possible to assume that orbital fluctuations cause the formation of heavy carriers in $Cd_2Re_2O_7$.

Table I Electronic parameters for phases I and III of Cd$_2$Re$_2$O$_7$

|  | $\rho_0$ (µΩcm) | $A$ ($10^{-2}$ µΩcm/K$^2$) | $n$ ($10^{20}$ cm$^{-3}$) | $\gamma_{exp}/\gamma_{band}$ (mJ/K$^2$ mol Re) | $\chi_s$ ($10^{-4}$ cm$^3$/mol Re) |
|---|---|---|---|---|---|
| Phase I | 247.4 | 4.5 | 1.47$^a$ | 30 ~ 50/5.73 | 1.65 |
| Phase III | 26.7 (11.5)$^b$ | 0.40 | 8.1 | 15.1/- | 0.75 |

$^a$From the band structure calculations.
$^b$The lowest value thus far obtained.

For the high-temperature phases, no specific heat data is available to directly determine the $\gamma$ value. However, we can estimate it from the susceptibility or resistivity data. As already reported, the change at $T_{s2}$ is relatively small, suggesting similar DOS for phases II and III. On the other hand, $\chi_s$ increases rapidly on heating near $T_{s1}$, and takes a value of 1.65 x $10^{-4}$ cm$^3$/mol Re above 200 K.[1, 5] It has been found that this large enhancement is due to the increase of DOS.[5] Therefore, assuming that $\chi_s$ is proportional to $\gamma$, we obtained $\gamma$ = 33 mJ/K$^2$ mol Re for phase I. In an alternative method, assuming that the Kadowaki-Woods ratio is the same for phases I and III leads us to a $\gamma$ value of 51 mJ/K$^2$ mol Re. It is likely that the mass enhancement in phase I is much larger than that in phase III. The band structure calculations using $x$ = 0.31728 gave $\gamma$ = 5.73 mJ/K$^2$ mol Re for phase I. Therefore, a large mass enhancement by a factor 5 ~ 8 may be realized in phase I. As already mentioned, the degeneracy of the hole bands must be larger in phase I than in phase II. It is plausible that orbital fluctuations are enhanced in phase I due to this large band degeneracy.

In conclusion, we have characterized the electronic properties of $Cd_2Re_2O_7$. The results obtained have indicated a semimetallic band structure with low carrier density for the high-temperature phase I with the ideal pyrochlore structure, in good agreement with the band structure calculations. It is found, moreover, that low-temperature phase III with a slightly distorted structure is also a semimetal with significantly smaller DOS and larger carrier density. A large mass enhancement is revealed for the two phases, particularly for phase I.

Most metallic pyrochlore oxides in the absence of structural deformations are poor metals. In addition, an intermetallic compound $YInCu_4$, which constitutes the phyrochlore lattice of Cu, is a semimetal showing a peculiar temperature dependence of resistivity.[16] It seems that there is a common electronic feature among these pyrochlore systems. $Cd_2Re_2O_7$ exists near the boundary between poor and good metals and thus exhibits characteristic symmetry-breaking transitions as a function of temperature.

Acknowledgments

We are grateful to M. Takigawa and K. Miyake for helpful discussions. We thank A.W. Sleight for providing us with his unpublished specific heat data. This research was supported by a Grant-in-Aid for Scientific Research on Priority Areas (A) and a Grant-in-Aid for Creative Scientific Research provided by the Ministry of Education, Culture, Sports, Science and Technology, Japan.